\documentclass[10pt, prb, aps, twocolumn, showpacs, citeautoscript, floatfix, reprint, amsmath, amssymb, notitlepage, superscriptaddress]{revtex4-1}

\usepackage{graphicx}
\usepackage{stmaryrd}
\usepackage{rotating}
\usepackage{amsmath}
\usepackage{amsfonts}
\usepackage{amssymb}
\usepackage{wasysym}
\usepackage[countmax]{subfloat}
\usepackage{dcolumn} 
\usepackage{bm} 
\usepackage{color}

\usepackage{float}
\usepackage{latexsym}

\begin{document}

\title{Probing disorder-induced Fisher information matrix and Cram\'{e}r-Rao bound by STM}


\author{Lucas A. Oliveira}

\affiliation{Department of Physics, PUC-Rio, 22451-900 Rio de Janeiro, Brazil}

\author{Wei Chen}

\affiliation{Department of Physics, PUC-Rio, 22451-900 Rio de Janeiro, Brazil}

\date{\rm\today}

\begin{abstract}

The electronic local density of states of solids, if normalized correctly, represents the probability density that the electron at a specific position has a particular energy. Because this probability density can vary in space in disordered systems, we propose that one can either treat the energy as a random variable and position as an external parameter to construct a real space Fisher information matrix, or treat the position as a random variable and energy as an external parameter to construct an energy space Fisher information, both quantify the variation of local density of states caused by the disorder. The corresponding Cram\'{e}r-Rao bounds in these two scenarios set a limit on the energy variance and the position variance of electrons, respectively, pointing to new interpretations of STM measurements. Our formalism thus bring the notion of information geometry into STM measurements, as demonstrated explicitly by lattice models of metals and topological insulators.

\end{abstract}

\maketitle

\section{Introduction}

The presence of disorder can significantly modify the local and global electronic properties of solids, offering the possibility of engineering these properties by disorder, as has been practiced experimentally and commercially in semiconductors for decades\cite{Bourgoin83,Queisser98,Alkauskas16}. Focusing on the local properties alone, various electronic properties are known to respond to the impurities in a spatially dependent manner, such as the Friedel oscillations of electron density near impurities in metals\cite{Friedel52}. From this perspective, the invention of scanning tunneling microscope (STM) brings the experimental detection of impurity effects to a whole new level, since it enables the detection of the spatial and energy dependence of electronic local density of states (LDOS) down to atomic scale\cite{Binnig82,Binnig82_2,Binnig87,Hansma87,Kubby96}. From this LDOS, various impurity effects have been experimentally verified, such as the impurity bound states\cite{Yazdani99,Balatsky06} and quasiparticle interference\cite{Hoffman02} in cuprate superconductors, among countless many others, offering a direct experimental confirmation of various theoretical predictions.

In this paper, we connect the disorder-induced LDOS to an arena that has not been explored much in condensed matter physics, but has been heavily exploited in statistics, namely the information geometry. We observe that if one combines the spatial and energy dependence of LDOS with the Born interpretation of quantum mechanical wave function, which states that correctly normalized LDOS $\rho({\bf x},E)$ is the probability density of finding an electron at a specific position ${\bf x}=(x^{1},x^{2}...x^{D})$ in a $D$-dimensional lattice with a particular energy $E$, then the notion of classical Fisher information matrix (FIM) naturally arise\cite{Fisher25}. Statistically, the notion of FIM arises when one has a probability distribution over some random variable whose profile depends on certain external parameters, where the FIM characterizes the variation of the probability distribution in the parameter space. Because of the probability interpretation of LDOS, one can adopt two scenarios to construct an FIM from the standard procedure in the probability theory. In the first scenario, one treats the energy $E$ as a random variable and the position ${\bf x}$ as an external parameter, i.e., the normalized LDOS $\rho({\bf x},E)$ is regarded as the probability of finding an electron within energy window $E$ and $E+\Delta E$ at a specified position ${\bf x}$, which allows to introduce a real space FIM to describe the variation of LDOS in real space caused by the disorder. In the second scenario, one reverses the roles by treating the position ${\bf x}$ as a random variable and the energy $E$ as an external parameter, meaning that the normalized LDOS $\rho({\bf x},E)$ is regarded as the probability of finding an electron at position ${\bf x}$ with a specified energy $E$, which amounts to an energy space FIM that also quantifies the variation of LDOS. The real space FIM and the energy space FIM, respectively, describe the distortion of the $D$-dimensional real space and the one-dimensional energy space caused by the disorder, thereby bringing the notion of information geometry\cite{Amari00,Amari16} into condensed matter physics. Furthermore, note that FIM has been applied to various branches of condensed matter physics, such as density functional theory\cite{Nagy22}, phase transitions\cite{Prokopenko11}, diamagnetism\cite{Curilef05}, etc. Our formalism thus demonstrates the LDOS as another application of FIM in a way that is completely measurable by STM.



In statistics, an important consequence of the FIM is the Cram\'{e}r-Rao bound (CRB), which sets a limit on the accuracy of parameter estimation\cite{Cramer46,Rao45}. For the real space FIM, we will elaborate that the corresponding CRB has a remarkable interpretation in electronic systems, namely it dictates that the variance of local energy $\langle E^{2}\rangle-\langle E\rangle^{2}$ must be greater than a product of the spatial derivative of the local energy $\partial_{\mu}\langle E\rangle$ and the inverse of the FIM, regardless the detail of the host material and what kind of disorder is present. In comparison, the energy space FIM gives a CRB that limits the spatial variance of electron position $\langle(x^{\mu})\rangle^{2}-\langle x^{\mu}\rangle^{2}$ according to the energy derivative of the average position $\partial_{E}\langle x^{\mu}\rangle$ and the FIM, also applicable to any kind of material and disorder. These bonuds on the energy variance and position variance of electrons point to new interpretations and data analysis of LDOS that can be broadly applied to any STM measurements, as will be demonstrated explicitly using theoretical models of metals and topological insulators.


\section{Real space Fisher information matrix and Cram\'{e}r-Rao bound}

\subsection{Real space Fisher information matrix defined from local density of states \label{sec:real_space_FIM}}

\begin{figure}[ht]
\begin{center}
\includegraphics[clip=true,width=0.7\columnwidth]{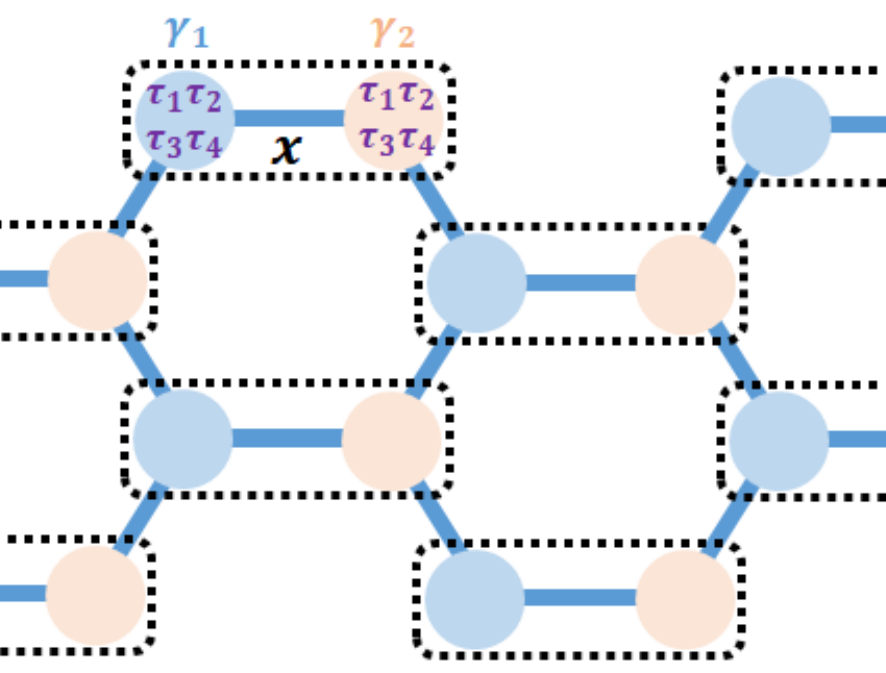}
\caption{Schematics of the notation in this paper using hexagonal lattice as an example, where we denote all the degrees of freedom in a unit cell as $\sigma=\gamma\otimes\tau$, where $\gamma$ is the sublattice index, and $\tau$ labels all the degrees of freedom on a lattice site such as spins and orbitals. The position of the unit cell is denoted by ${\bf x}$, and the LDOS measured by STM at sublattice $\gamma$ at position ${\bf x}$ is $\tilde{\rho}_{\gamma}({\bf x},E)$. } 
\label{fig:sigma_gamma_x_schematics}
\end{center}
\end{figure}

We consider a $D$-dimensional lattice with unit vectors ${\bf a}^{\mu}=a^{\mu}{\hat{\boldsymbol\mu}}$, where $\mu=1\sim D$ is reserved for the spatial directions. The locations of the unit cell, or lattice sites, are described by the Bravais lattice vector ${\bf x}=\sum_{\mu=1}^{D}n_{\mu}{\bf a}^{\mu}$, where $n_{\mu}$ are integers. Each unit cell may have several species of electrons labeled by $\sigma$, which may signify spin, orbit, sublattice, etc. Generically, the lattice Hamiltonian takes the form
\begin{eqnarray}
H=\sum_{\bf xx'\sigma\sigma '}t_{\bf xx'\sigma\sigma '}|{\bf x},\sigma\rangle\langle{\bf x}',\sigma '|,
\end{eqnarray}
where $|{\bf x},\sigma\rangle$ is the ket state of an electron of species $\sigma$ at site ${\bf x}$. The lattice eigenstates $|E_{\ell}\rangle$ and eigenenergies $E_{\ell}$ are found through diagonalizing the Hamiltonian $H|E_{\ell}\rangle=E_{\ell}|E_{\ell}\rangle$.
The wave function of species $\sigma$ at site ${\bf x}$ of an eigenstate $|E_{\ell}\rangle$ is denoted by
\begin{eqnarray}
&&\langle {\bf x},\sigma|E_{\ell}\rangle=\tilde{a}_{\ell\sigma}({\bf x}).
\label{wavefn_atilde_nisigma}
\end{eqnarray}
Throughout the article, we reserve the subscript $n$ for the filled states at zero temperature within the energy range of an STM measurement $E_{c}<E_{n}<0$, where $E_{c}$ denotes the lowest energy measured by STM (typically few eV), and likewisely for the summation $\sum_{n}\equiv\sum_{E_{c}<E_{n}<0}$. Certainly STM also measures positive energy states $E_{\ell}>0$ up to some highest energy, but they are unimportant for our discussion since the electrons at zero temperature do not occupy these states. Furthermore, usually STM do not measure all the degrees of freedoms $\sigma$ within a unit cell separately. For instance, the two spins are usually not resolved by STM. We will assume that the STM can resolve the LDOS on each sublattice $\gamma$ within a unit cell, but cannot resolve any other degrees of freedom $\tau$ such as spins and orbitals, and hence the measured LDOS effectively sums over $\tau$. For instance, for a system that contains two sublattices $\gamma=\left\{A,B\right\}$, and two orbitals and two spins $\tau=\left\{s\uparrow,s\downarrow,p\uparrow,p\downarrow\right\}$, the label for degrees of freedom $\sigma$ can be split into
\begin{eqnarray}
\sigma=\left\{A,B\right\}\otimes\left\{s\uparrow,s\downarrow,p\uparrow,p\downarrow\right\}
=\gamma\otimes\tau,
\label{sigma_gamma_tau}
\end{eqnarray}
as shown schematically in Fig.~\ref{fig:sigma_gamma_x_schematics}. Having these notations defined, we are ready to discuss the STM measurements.

The precise theoretical description of the tunneling process in the STM measurements involves a fair amount of complications, such as the shape and atomic wave function of the tip, transfer matrix elements, etc\cite{Bardeen61,Tersoff83,Tersoff85,Chen90,Chen90_2,Chen21_STM_book}. Nevertheless, on a phenomenological level, one may assume that the STM measures the tunneling conductance in the energy range $E_{c}<E<0$ on the sublattice $\gamma$ of a unit cell located at ${\bf x}$ described by\cite{Hoffman02} 
\begin{eqnarray}
\frac{dI}{dV}({\bf x},\gamma,E)=\frac{4\pi e}{\hbar}|M|^{2}\tilde{\rho}_{\rm tip}(0)\,\tilde{\rho}_{\gamma}({\bf x},E),
\label{dIdV_general_formula}
\end{eqnarray}
where the tunneling matrix element $|M|^{2}$ and the DOS of the tip $\tilde{\rho}_{\rm tip}(0)$ are assumed to be independent of energy $E$, and $\tilde{\rho}_{\gamma}({\bf x},E)$ is the LDOS on the sublattice $\gamma$ of the unit cell at ${\bf x}$. The LDOS $\tilde{\rho}_{\gamma}({\bf x},E)$ is calculated from the wave function in Eq.~(\ref{wavefn_atilde_nisigma}) in the following manner. For each degree of freedom $\sigma=\gamma\otimes\tau$, the LDOS within the probed range is given by 
\begin{eqnarray}
\tilde{\rho}_{\gamma\otimes\tau}({\bf x},E)=\sum_{n}|\tilde{a}_{n\gamma\otimes\tau}({\bf x})|^{2}\delta(E-E_{n}).
\end{eqnarray}
But since STM cannot resolve $\tau$ such as spins and orbitals, the measured LDOS on sublattice $\gamma$ contains a summation over $\tau$
\begin{eqnarray}
&&\tilde{\rho}_{\gamma}({\bf x},E)=\sum_{\tau}\tilde{\rho}_{\gamma\otimes\tau}({\bf x},E)
\nonumber \\
&&=\sum_{n}\left(\sum_{\tau}|\tilde{a}_{n\gamma\otimes\tau}({\bf x})|^{2}\right)\delta(E-E_{n})
\nonumber \\
&&\equiv\sum_{n}|\tilde{a}_{n\gamma}({\bf x})|^{2}\delta(E-E_{n})
\label{tilderho_an_expression}
\end{eqnarray}
which gives the differential conductance measured by STM in Eq.~(\ref{dIdV_general_formula}). For the sake of formulating an FIM, we proceed to consider the summation of the LDOS over all the sublattices $\gamma$ at ${\bf x}$
\begin{eqnarray}
&&\tilde{\rho}({\bf x},E)=\sum_{\gamma}\tilde{\rho}_{\gamma}({\bf x},E)=\sum_{\gamma\tau}\tilde{\rho}_{\gamma\otimes\tau}({\bf x},E)
\nonumber \\
&&=\sum_{n}\left(\sum_{\gamma\tau}|\tilde{a}_{n\gamma\otimes\tau}({\bf x})|^{2}\right)\delta(E-E_{n})
\nonumber \\
&&\equiv\sum_{n}|\tilde{a}_{n}({\bf x})|^{2}\delta(E-E_{n}).
\label{tilderho_sum_gamma}
\end{eqnarray}
which defines the amplitude $|\tilde{a}_{n}({\bf x})|^{2}$ at position ${\bf x}$.

We proceed to define a normalized LDOS by dividing the $\tilde{\rho}({\bf x},E)$ in Eq.~(\ref{tilderho_sum_gamma}) by its energy integration
\begin{eqnarray}
&&\rho({\bf x},E)=\frac{\tilde{\rho}({\bf x},E)}{\int_{E_{c}}^{0}dE\,\tilde{\rho}({\bf x},E)}=\frac{\tilde{\rho}({\bf x},E)}{n({\bf x},E_{c})}
\nonumber \\
&&=\sum_{n}\frac{|\tilde{a}_{n}({\bf x})|^{2}}{n({\bf x},E_{c})}\delta(E-E_{n})
\equiv\sum_{n}|a_{n}({\bf x})|^{2}\delta(E-E_{n}),\;\;\;
\label{rhoxE_normalized}
\end{eqnarray}
such that it is normalized correctly
\begin{eqnarray}
&&1=\int_{E_{c}}^{0}dE\,\rho({\bf x},E)=\sum_{n}|a_{n}({\bf x})|^{2}.
\end{eqnarray}
where $n({\bf x},E_{c})$ is simply the number of electrons resided at ${\bf x}$ (counting all the sublattices $\gamma$ and spins and orbitals $\tau$) with energy $E_{c}<E<0$. This normalization allows us to treat the normalized LDOS $\rho({\bf x},E)$ as a conditional probability density function (PDF), since it is the probability density that an electron at position ${\bf x}$ has energy $E$ under the condition $E_{c}<E<0$ set by the STM probe range. 


Experimentally, one can simply extract $\rho({\bf x},E)$ from the tunneling conductance in Eq.~(\ref{dIdV_general_formula}) summing over all sublattices $\gamma$ at ${\bf x}$ divided by its energy integration 
\begin{eqnarray}
&&\sum_{\gamma}\frac{dI}{dV}({\bf x},\gamma,E)\bigg/\sum_{\gamma}\int_{E_{c}}^{0}dE\frac{dI}{dV}({\bf x},\gamma,E)
\nonumber \\
&&=\rho({\bf x},E),
\end{eqnarray}
where we see that all other complications such as tip DOS $\tilde{\rho}_{\rm tip}(0)$ and tunneling matrix element $|M|^{2}$ have been divided out under the assumption that they do not depend on energy $E$, which serves as a concrete experimental protocol to extract the $\rho({\bf x},E)$. Having this PDF defined and measured, we proceed to introduce its FIM from the usual definition in the probability theory\cite{Fisher25}
\begin{eqnarray}
&&I_{\mu\nu}({\bf x},E_{c})=\int_{E_{c}}^{0}dE\,\frac{\partial_{\mu}\rho({\bf x},E)\partial_{\nu}\rho({\bf x},E)}{\rho({\bf x},E)},
\label{FIM_extracted_from_STM}
\end{eqnarray}
which can be derived from the leading order expansion of the fidelity between two normalized LDOSs that are very close in real space
\begin{eqnarray}
\int_{E_{c}}^{0}dE\sqrt{\rho({\bf x},E)}\sqrt{\rho({\bf x+\delta x},E)}=1-\frac{1}{8}I_{\mu\nu}\delta x^{\mu}\delta x^{\nu}.\;\;\;\;\;\;
\end{eqnarray}
This expansion is very similar to the recently proposed real space quantum metric derived from the overlap between two neighboring local states constructed from projecting the ground state to lattice sites\cite{Oliveira25_real_space_quantum_metric,Provost80}. Within the context of information geometry\cite{Amari00,Amari16}, this FIM acts like a metric on the Euclidean manifold of real space ${\bf x}$ that describes how the PDF $\rho({\bf x},E)$ varies with ${\bf x}$, and it depends on the energy range $E_{c}$ measured in an STM experiment.

We proceed to generalize the FIM and quantum metric to all different types of crystalline structures by the following argument: Consider two different types of lattices, say monoclinic and orthorhombic. If they are both described by the same hopping $\left\{t_{a},t_{b},t_{c}\right\}$ on the edges, then their eigenstates $|E_{\ell}\rangle$ and wave functions $\tilde{a}_{\ell\sigma}({\bf x})$ will be exactly the same, just that their wave functions will live on lattice sites ${\bf x}=\sum_{\mu=1}^{D}n_{\mu}{\bf a}^{\mu}$ that are located at different points in space because of the two different crystalline structures. Although the Bravais unit lattice vectors ${\bf a}^{\mu}=a^{\mu}{\hat{\boldsymbol\mu}}$ are not necessarily orthogonal to each other in certain crystalline structures, such as in monoclinic lattices, the above argument suggests that we should still define the derivatives along these lattice vectors, because only the network connectivity of the lattice points matter for the eigenstates and eigenvectors. For the lattices with a basis, such as the hexagonal lattice shown in Fig.~\ref{fig:sigma_gamma_x_schematics}, the index $\gamma$ takes care of the sublattices. In practice, we calculate the derivatives in Eq.~(\ref{FIM_extracted_from_STM}) along the lattice vectors numerically by the central difference\cite{Oliveira25_real_space_quantum_metric}
\begin{eqnarray}
\partial_{\mu}\rho({\bf x},E)=\frac{1}{2a^{\mu}}\left[\rho({\bf x}+{\bf a}^{\mu},E)-\rho({\bf x}-{\bf a}^{\mu},E)\right],
\nonumber \\
\label{spatial_derivative_numerical}
\end{eqnarray}
instead of the forward difference $\left[\rho({\bf x}+{\bf a}^{\mu},E)-\rho({\bf x},E)\right]/a^{\mu}$ or backward difference $\left[\rho({\bf x},E)-\rho({\bf x}-{\bf a}^{\mu},E)\right]/a^{\mu}$. This is based on the reasoning that the FIM as a metric should be symmetric under the inversion of the small displacement $\delta{\bf x}={\bf a}^{\mu}$ to $\delta{\bf x}=-{\bf a}^{\mu}$, and moreover the method will result in a FIM $I_{\mu\nu}$ that are spatially symmetric around a single impurity.



\subsection{Cram\'{e}r-Rao bound on the energy variance of electrons \label{sec:CRB_real_space}}

A profound application of the FIM is the CRB. Traditionally, the CRB sets a limit on the accuracy of an estimator in the problem of parameter estimation. However, this interpretation is not very relevant to STM, since the experimentalist presumably knows all the parameters in the measurement, such as the position of the tip ${\bf x}$ and the bias energy $E$. Instead, below we elaborate that the CRB can be interpreted as a lower bound on the variance of  physical measurables. Treating the normalized LDOS $\rho({\bf x},E)$ in Eq.~(\ref{rhoxE_normalized}) as a PDF, we follow the formalism of van den Bos\cite{vandenBos07} to derive a bound on the variance of any function of energy. Consider any arbitrary function of energy $f(E)$ whose expected value at position ${\bf x}$ is
\begin{eqnarray}
\langle f(E)\rangle=\int dE\,\rho({\bf x},E)f(E)\equiv\langle f\rangle.
\label{estimator_bias_definition}
\end{eqnarray}
where $\langle...\rangle$ denotes the expected value calculated with respect to the PDF $\rho({\bf x},E)$. The resulting $\langle f\rangle$ is a function of position ${\bf x}$, whose spatial derivative $\partial_{\mu}=\partial/\partial x^{\mu}$ satisfies
\begin{eqnarray}
&&\partial_{\mu}\langle f\rangle=\int dE\,f(E)\,\partial_{\mu}\rho=\int dE\left(f-\langle f\rangle\right)\rho\,\partial_{\mu}\ln\rho
\nonumber \\
&&=\langle\left(f-\langle f\rangle\right)\partial_{\mu}\ln\rho\rangle,
\label{df_formula}
\end{eqnarray}
since $\langle\partial_{\mu}\ln\rho\rangle=0$. 

We proceed to introduce a vector of $D+1$ elements 
\begin{eqnarray}
{\bf m}=\left(\begin{array}{c}
f(E)-\langle f\rangle \\
\partial_{\nu}\ln\rho({\bf x},E)
\end{array}\right)\equiv\left(\begin{array}{c}
B^{1} \\
\vdots \\
B^{D+1}
\end{array}\right).
\label{m_vector}
\end{eqnarray}
The expected value of the outer product of this vector with itself is
\begin{eqnarray}
&&\langle{\bf m}\otimes{\bf m}^{T}\rangle
=\left(\begin{array}{cc}
\langle(f-\langle f\rangle)^2\rangle & \langle(f-\langle f\rangle)\partial_{\eta}\ln\rho\rangle \\
\langle\partial_{\nu}\ln\rho(f-\langle f\rangle)\rangle & I_{\nu\eta}
\end{array}\right),
\nonumber \\
\end{eqnarray} 
where $I_{\nu\eta}$ is the FIM. For any real vector ${\bf v}^{T}=(v_{1}...v_{D+1})$, the product with the above matrix gives
\begin{eqnarray}
{\bf v}^{T}\cdot\langle{\bf m}\otimes{\bf m}^{T}\rangle\cdot{\bf v}
=\int dE\,\rho\left(v_{\mu}B^{\mu}\right)^{2}\geq 0,\;\;\;
\end{eqnarray}
after using Eq.~(\ref{m_vector}), indicating that $\langle{\bf m}\otimes{\bf m}^{T}\rangle$ is positive semidefinite. Choosing ${\bf v}^{T}=\left(1,-\partial\langle f\rangle^{T}I^{-1}\right)$ and combining with Eq.~(\ref{df_formula}), this positive semidefiniteness of $\langle{\bf m}\otimes{\bf m}^{T}\rangle$ gives the following multiplication of matrices
\begin{eqnarray}
&&\left(1,-\partial\langle f\rangle^{T}I^{-1}\right)
\left(\begin{array}{cc}
\langle f^2\rangle-\langle f\rangle^{2} & \partial\langle f\rangle^{T} \\
\partial\langle f\rangle & I
\end{array}\right)
\left(\begin{array}{c}
1 \\
-I^{-1}\partial\langle f\rangle
\end{array}\right)
\nonumber \\
&&=\langle f^2\rangle-\langle f\rangle^{2}
-\partial\langle f\rangle^{T}I^{-1}\partial\langle f\rangle\geq 0.
\end{eqnarray}
Rearranging the above inequality and writing down the indices explicitly give the CRB (repeating indices are summed)
\begin{eqnarray}
\langle f^2\rangle-\langle f\rangle^{2}\geq 
\partial_{\rho}\langle f\rangle I^{\rho\lambda}\partial_{\lambda}\langle f\rangle
\nonumber \\
\label{CRB_space}
\end{eqnarray}
where $I^{\rho\lambda}$ is the inverse of FIM that satisfies $I^{\rho\lambda}I_{\lambda\xi}=\delta_{\xi}^{\rho}$. Both sides of Eq.~(\ref{CRB_space}) depend on position ${\bf x}$, meaning that this inequality must be satisfied at any ${\bf x}$. Note that the inverse of FIM $I^{\rho\lambda}$ does not exist for homogeneous or very weakly disordered systems where the FIM vanishes $I_{\lambda\xi}\rightarrow 0$ in all or some parts of the space, in which case the CRB in Eq.~(\ref{CRB_space}) is also ill-defined. In other words, the CRB in Eq.~(\ref{CRB_space}) is applicable everywhere in highly disordered systems where the FIM is nonzero throughout the systems, but in systems with dilute impurities it only applies to regions around the impurity sites.

The physical interpretation of Eq.~(\ref{CRB_space}) is fairly straightforward: It simply means that the variance of any energy function $f(E)$ at site ${\bf x}$ must be greater than the product of spatial derivative of its expected value $\partial_{\rho}\langle f\rangle$ and the inverse of FIM $I^{\rho\lambda}$. Note that the right hand side of Eq.~(\ref{CRB_space}) is coordinate invariant, i.e., under some unitary transformation $U$ that changes the coordinates to $\partial_{\mu}^{\prime}=U_{\mu}^{\rho}\partial_{\rho}$ without changing the underlying crystalline structure, such as inversion or rotation, the right hand side of Eq.~(\ref{CRB_space}) is unchanged, so one can calculate the derivative $\partial_{\rho}$ along any choice of crystalline directions.

A particularly important application of Eq.~(\ref{CRB_space}) is when we choose the energy function to be the energy itself $f(E)=E$. In this case, $\langle f\rangle=\langle E\rangle$ is the local energy of electrons resided at position ${\bf x}$, and $\langle f^{2}\rangle-\langle f\rangle^{2}=\langle E^{2}\rangle-\langle E\rangle^{2}$ is simply the variance of local energy, so the CRB simply sets a low bound on this local energy variance according to the spatial variation of the local energy $\partial_{\rho}\langle E\rangle$ and the inverse of FIM $I^{\rho\lambda}$. Explicitly for 1D systems, the CRB reads 
\begin{eqnarray}
\langle E^{2}\rangle-\langle E\rangle^{2}\geq\frac{\left(\partial_{\rho}\langle E\rangle\right)^{2}}{I_{xx}},
\end{eqnarray}
that can also be derived from the Cauchy-Schwarz inequality\cite{vandenBos07}. In 2D systems, the CRB is calculated by the matrix product
\begin{eqnarray}
\langle E^{2}\rangle-\langle E\rangle^{2}\geq\left(\partial_{x}\langle E\rangle\;\;\partial_{y}\langle E\rangle\right)\left(\begin{array}{cc}
I^{xx} & I^{xy} \\
I^{yx} & I^{yy}
\end{array}\right)\left(\begin{array}{c}
\partial_{x}\langle E\rangle \\
\partial_{y}\langle E\rangle
\end{array}\right),
\nonumber \\
\label{CRB_E_2D}
\end{eqnarray}
and likewisely for 3D systems. The 2D version of CRB is particularly useful for STM, since STM usually measures either a 2D material\cite{Andrei12} or the topmost 2D layer of a 3D material, so it has access to the 2D position ${\bf x}=(x,y)$ that are treated as a 2D parameter space, while it is impotent to measure the $z$ direction.

We remark that the local energy variance $\langle E^{2}\rangle-\langle E\rangle^{2}$ defined via the expected value in Eq.~(\ref{estimator_bias_definition}) is not the energy fluctuation that enters the fluctuation-dissipation theorem. Recall that the heat capacity $C_{V}$ of the system is related to energy fluctuation $\overline{H^{2}}-\overline{H}^{2}$ by
\begin{eqnarray}
C_{V}=\frac{1}{k_{B}T^{2}}\left(\overline{H^{2}}-\overline{H}^{2}\right),
\end{eqnarray}
where the ensemble average of the Hamiltonian to the $\alpha$-th power can be calculated from the lattice eigenstates by
\begin{eqnarray}
&&\overline{H^{\alpha}}=\sum_{n}f(E_{n})\langle E_{n}|H^{\alpha}|E_{n}\rangle
\nonumber \\
&&=\sum_{n}f(E_{n})\sum_{\bf x}\sum_{\gamma\tau}|\tilde{a}_{n\gamma\otimes\tau}({\bf x})|^{2}(E_{n})^{\alpha}.
\end{eqnarray}
with $f(E_{n})$ the Fermi distribution. Thus the energy fluctuation $\overline{H^{2}}-\overline{H}^{2}$ and the local energy variance we proposed $\langle E^{2}\rangle-\langle E\rangle^{2}$ are not the same object, and consequently the later does not determine the heat capacity $C_{V}$. Nevertheless, the local energy variance serves as a useful tool to characterize the influence of disorder on the local energy distribution, and moreover gives rise to the CRB introduced above, as we shall see in the following sections using concrete 2D models.

\begin{figure*}[ht]
\begin{center}
\includegraphics[clip=true,width=1.8\columnwidth]{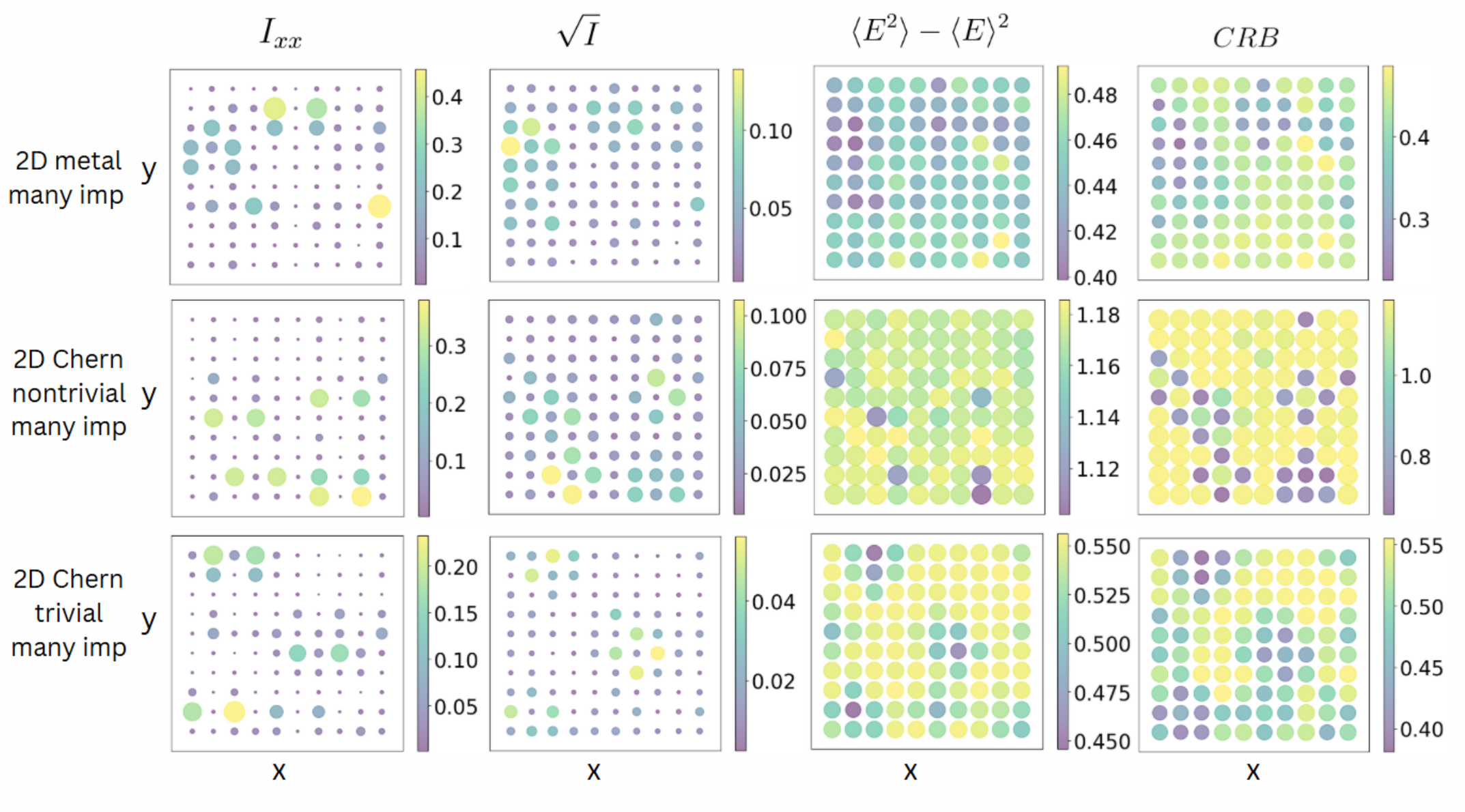}
\caption{Numerical results for the real space information geometry in disordered 2D metal (top), 2D Chern insulator in the topologically nontrivial phase (middle) and trivial phase (bottom), each calculated in one multiple-impurity configuration. For each system, we present the FIM along ${\bf x}$ direction $I_{xx}$, the volume form $\sqrt{I}=\sqrt{\det I_{\mu\nu}}$, the energy variance $\langle E^{2}\rangle-\langle E\rangle^{2}$, and the CRB defined as $\langle E^2\rangle-\langle E\rangle^{2}- 
\partial_{\rho}\langle E\rangle I^{\rho\lambda}\partial_{\lambda}\langle E\rangle$. All these quantities should be positive on every site regardless of the system parameters and impurity configuration. } 
\label{fig:real_space_CRB_figure}
\end{center}
\end{figure*}

\subsection{Applications to 2D materials \label{sec:applications_2D_materials}}

For concreteness, we use 2D tight-binding models of metals and Chern insulators to demonstrate the FIM and CRB. The 2D models are defined on a square lattice that has no sublattice, so we drop the subscript $\gamma$ for convenience, and we will assume that the STM measures the entire negative energy window such that $E_{c}=-\infty$. We first consider 2D spinless fermions with nearest-neighbor hopping and multiple potential impurities
\begin{eqnarray}
H&=&-t\sum_{\bf \langle xx'\rangle}\left(c_{\bf x}^{\dag}c_{\bf x'}+c_{\bf x'}^{\dag}c_{\bf x}\right)-t'\sum_{\bf \langle\langle xx''\rangle\rangle}\left(c_{\bf x}^{\dag}c_{\bf x''}+c_{\bf x''}^{\dag}c_{\bf x}\right)
\nonumber \\
&&-\mu\sum_{\bf x}c_{\bf x}^{\dag}c_{\bf x}+U_{imp}\sum_{{\bf x}\in imp}c_{\bf x}^{\dag}c_{\bf x},
\end{eqnarray}
where ${\bf x}=(x,y)$ labels the position of the lattice sites, $t=1$ is the nearest-neighbor hopping, $t'=0.5$ is the next-nearest-neighbor hopping, $\mu=0.1$ is the chemical potential, and the impurity potential takes random values within the range $0<U_{imp}<1$ on $10\%$ of the sites. Since the model is spinless and single orbital, we drop the index $\tau$ as well.

Secondly, we examine the lattice model of Chern insulators with multiple impurities described by\cite{Chen20_absence_edge_current,Molignini23_Chern_marker} 
\begin{eqnarray}
&&H=\sum_{\bf x}t\left\{-ic_{{\bf x}s}^{\dag}c_{{\bf x+a}^{x}p}
+ic_{{\bf x+a}^{x}s}^{\dag}c_{{\bf x}p}+h.c.\right\}
\nonumber \\
&&+\sum_{\bf x}t\left\{-c_{{\bf x}s}^{\dag}c_{{\bf x+a}^{y}p}+c_{{\bf x+a}^{y}s}^{\dag}c_{{\bf x}p}+h.c.\right\}
\nonumber \\
&&+\sum_{{\bf x}}\sum_{\delta=\left\{x,y\right\}}t'\left\{-c_{{\bf x}s}^{\dag}c_{{\bf x+a}^{\delta} s}+c_{{\bf x}p}^{\dag}c_{{\bf x+a}^{\delta} p}+h.c.\right\}
\nonumber \\
&&+\sum_{\bf x}\left(M+4t'\right)\left\{c_{{\bf x}s}^{\dag}c_{{\bf x}s}
-c_{{\bf x}p}^{\dag}c_{{\bf x}p}\right\},
\nonumber \\
&&-\mu\sum_{\bf x}\left\{c_{{\bf x}s}^{\dag}c_{{\bf x}s}
+c_{{\bf x}p}^{\dag}c_{{\bf x}p}\right\},
\nonumber \\
&&+U_{imp}\sum_{{\bf x}\in imp}\left\{c_{{\bf x}s}^{\dag}c_{{\bf x}s}
+c_{{\bf x}p}^{\dag}c_{{\bf x}p}\right\},
\label{Hamiltonian_2DclassA}
\end{eqnarray} 
where $t=1$ and $t'=0.5$ are the nearest-neighbor hopping between the opposite and the same orbitals, respectively, $\mu=0.1$ is the chemical potential, ${\bf a}=(a^{x},a^{y})$ are the unit vectors in the two planar directions, and $M$ is the mass term that controls the topological order. The impurity potential $U_{imp}$ is the same for the two orbitals $\tau=\left\{s,p\right\}$ on $10\%$ of sites, and takes random values within the interval $0<U_{imp}<1$. The effects of multiple impurities and different kinds of impurity potential have been investigated intensively in this model, especially on the issue of conservation of topological order investigated by means of the Chern marker\cite{Bianco11,Prodan10,Costa19,Ulcakar20,dOrnellas22,Oliveira24_impurity_marker}.

Numerical results for the metal, Chern insulator in the topologically nontrivial phase $M=-2$ and trivial phase $M=2$ are shown in Fig.~\ref{fig:real_space_CRB_figure}, each calculated for one single realization of impurity configuration. For each case, we calculate the FIM $I_{xx}$, volume form $\sqrt{I}=\sqrt{\det I_{\mu\nu}}$, energy variance $\langle E^{2}\rangle-\langle E\rangle^{2}$, and the CRB defined as $\langle E^2\rangle-\langle E\rangle^{2}- 
\partial_{\rho}\langle E\rangle I^{\rho\lambda}\partial_{\lambda}\langle E\rangle$, and plot them as functions of position ${\bf x}=(x,y)$. All these quantities are positive on every site, in accordance with our theory. In particular, we see that impurities induce nonzero $I_{\mu\nu}$ and the volume form $\sqrt{I}$, meaning that the real space is curved by the presence of impurities in the sense of information geometry. Comparing the two results for the Chern insulator, the variance $\langle E^2\rangle-\langle E\rangle^{2}$ and CRB are found to be larger in the topologically nontrivial phase. The positive CRB implies that the energy variance $\langle E^2\rangle-\langle E\rangle^{2}$ is bounded by $\partial_{\rho}\langle E\rangle I^{\rho\lambda}\partial_{\lambda}\langle E\rangle$ for any kind of material and disorder, thereby giving an explicit example of how the FIM constraints the variance of energy functions.


\section{Energy space Fisher information matrix and Cram\'{e}r-Rao bound}

\subsection{Energy space Fisher information defined from local density of states}

In this section, we propose yet another interpretation of FIM by treating the position ${\bf x}$ as a random variable and energy $E$ as an external parameter. Because the energy is a 1D parameter, the Fisher information in this parameter space is just a number instead of a matrix, but we will still call it FIM for simplicity. Starting from the tunneling conductance measured by STM given in Eq.~(\ref{dIdV_general_formula}), we now consider an STM measurement over some finite region of space $0\leq x^{\mu}<x_{0}^{\mu}$. For instance, at a given energy $E\sim 10$meV, a typical STM image covers a region of $\sim 10$nm$\times 10$nm of the surface of a 2D or 3D sample. Summing the tunneling conductance over all the unit cells ${\bf x}$ and sublattices $\gamma$ over this finite region at a fixed energy $E$ yields 
\begin{eqnarray}
&&\frac{dI}{dV}(E)\equiv\sum_{\bf x\gamma}\frac{dI}{dV}({\bf x},\gamma,E)
\nonumber \\
&&=\frac{4\pi e}{\hbar}|M|^{2}\tilde{\rho}_{\rm tip}(0)\sum_{\bf x\gamma}\tilde{\rho}({\bf x},E)
\nonumber \\
&&=\frac{4\pi e}{\hbar}|M|^{2}\tilde{\rho}_{\rm tip}(0)\,\tilde{\rho}(E),
\label{dIdV_sumx_definition}
\end{eqnarray}
where $\tilde{\rho}(E)=\sum_{n}\delta(E-E_{n})$ is the density of states of the measured region. At a fixed unit cell position ${\bf x}$ and energy $E$, we sum the tunneling conductance over sublattices $\gamma$ and divide the above quantity to obtain
\begin{eqnarray}
&&\sum_{\gamma}\frac{dI}{dV}({\bf x},\gamma,E)\bigg/\frac{dI}{dV}(E)=\frac{\sum_{\gamma}\tilde{\rho}_{\gamma}({\bf x},E)}{\tilde{\rho}(E)}=\overline{\rho}({\bf x},E),
\nonumber \\
&&\sum_{\bf x}\overline{\rho}({\bf x},E)=1,
\end{eqnarray}
and hence $\overline{\rho}({\bf x},E)$ can be treated as a probability mass function (PMF) over the discrete variable ${\bf x}$, and the energy $E$ plays the role of an external parameter. Again we see that all the other complications of instrumentation, such as the tunneling matrix element $|M|^{2}$ and the tip DOS $\tilde{\rho}_{\rm tip}(0)$ have been divided out, so they do not affect the argument below. Because $\overline{\rho}({\bf x},E)$ can be treated as the probability of finding an electron at ${\bf x}$ with energy $E$, one can proceed to introduce an energy space FIM by
\begin{eqnarray}
I_{E}(E)=\sum_{\bf x}\frac{\left[\partial_{E}\overline{\rho}({\bf x},E)\right]^{2}}{\overline{\rho}({\bf x},E)},
\end{eqnarray}
which characterizes the fidelity between two normalized LDOSs that are very close in the energy space
\begin{eqnarray}
\sum_{\bf x}\sqrt{\overline{\rho}({\bf x},E)}\sqrt{\overline{\rho}({\bf x},E+\delta E)}=1-\frac{1}{8}I_{E}\,\delta E^{2}.
\end{eqnarray}
Numerically, the PMF and its derivative in the expression of FIM can be calculated from the $|\tilde{a}_{n\gamma}({\bf x})|^{2}$ defined in Eq.~(\ref{tilderho_an_expression}) and eigenenergy $E_{n}$ by
\begin{eqnarray}
&&\overline{\rho}({\bf x},E)=\sum_{n}|\tilde{a}_{n\gamma}({\bf x})|^{2}\left[\frac{\delta(E-E_{n})}{\sum_{n'}\delta(E-E_{n'})}\right],
\nonumber \\
&&\partial_{E}\overline{\rho}({\bf x},E)=\sum_{n}|\tilde{a}_{n\gamma}({\bf x})|^{2}\partial_{E}\left[\frac{\delta(E-E_{n})}{\sum_{n'}\delta(E-E_{n'})}\right]
\nonumber \\
&&=\sum_{n}|\tilde{a}_{n\gamma}({\bf x})|^{2}\left\{\frac{\partial_{E}\delta(E-E_{n})}{\sum_{n'}\delta(E-E_{n'})}\right.
\nonumber \\
&&\left.-\frac{\delta(E-E_{n})\sum_{n''}\partial_{E}\delta(E-E_{n''})}{\left[\sum_{n'}\delta(E-E_{n'})\right]^{2}}\right\},
\label{rho_dErho_expressions}
\end{eqnarray}
where the $\delta$-function and its derivative are interpreted by a Lorentzian
\begin{eqnarray}
&&\delta(E-E_{n})=\frac{\eta/\pi}{(E-E_{n})^{2}+\eta^{2}},
\nonumber \\
&&\partial_{E}\delta(E-E_{n})=-\frac{2(E-E_{n})\,\eta/\pi}{\left[(E-E_{n})^{2}+\eta^{2}\right]^{2}},
\label{delta_fn_dE_interpretation}
\end{eqnarray}
with an appropriate choice of artificial broadening $\eta$.

\subsection{Cram\'{e}r-Rao bound on the spatial variance of electrons \label{sec:CRB_energy_space}}

The construction of CRB for the energy space FIM $I_{E}$ is completely analogous to that in Sec.\ref{sec:CRB_real_space}, except one swaps the roles of $({\bf x},\gamma)$ and $E$. Consider any function $f({\bf x})$ of position ${\bf x}$, whose expected value at energy $E$ within the spatial region measured by STM is
\begin{eqnarray}
\langle f({\bf x})\rangle=\sum_{\bf x}f({\bf x})\overline{\rho}({\bf x},E)=\langle f\rangle, 
\end{eqnarray}
and hence $\langle f\rangle$ depends on energy $E$. Note that the bracket $\langle...\rangle$ here has a different meaning than that in Eq.~(\ref{estimator_bias_definition}), but we should stick to this notation that is frequently used in statistics. The energy derivative of this expected value is 
\begin{eqnarray}
\partial_{E}\langle f\rangle=\sum_{\bf x}f\partial_{E}\overline{\rho}
=\langle\left(f-\langle f\rangle\right)\partial_{E}\ln\overline{\rho}\rangle,
\end{eqnarray}
again due to $\langle\partial_{E}\ln\overline{\rho}\rangle=0$. The application of the Cauchy-Schwarz inequality to the square of this energy derivative gives\cite{vandenBos07}
\begin{eqnarray}
&&\left(\partial_{E}\langle f\rangle\right)^{2}=\left[\sum_{\bf x}\left(f-\langle f\rangle\right)\overline{\rho}\,\partial_{E}\ln\overline{\rho}\right]^{2}
\nonumber \\
&&\leq \left[\sum_{\bf x}\left(f-\langle f\rangle\right)^{2}\overline{\rho}\right]\cdot\left[\sum_{\bf x}\overline{\rho}\left(\partial_{E}\ln\overline{\rho}\right)^{2}\right],
\end{eqnarray}
which can be rearranged into a CRB
\begin{eqnarray}
\langle f^{2}\rangle-\langle f\rangle^{2}\geq\frac{\left(\partial_{E}\langle f\rangle\right)^{2}}{I_{E}},
\label{CRB_E_space}
\end{eqnarray}
that should be satisfied at any energy $E$. Numerically, the $\langle f^{2}\rangle$, $\langle f\rangle^{2}$, and $\partial_{E}\langle f\rangle$ in this expression can be calculated by using the expressions for $\overline{\rho}$ and $\partial_{E}\overline{\rho}$ in Eqs.~(\ref{rho_dErho_expressions}) and (\ref{delta_fn_dE_interpretation}), so the validity of Eq.~(\ref{CRB_E_space}) can be easily verified.

A particularly significant application of this CRB is when we take the function of position to be the position itself $f({\bf x})=x^{\mu}$. In this case, Eq.~(\ref{CRB_E_space}) implies
\begin{eqnarray}
\langle (x^{\mu})^{2}\rangle-\langle x^{\mu}\rangle^{2}\geq\frac{\left(\partial_{E}\langle x^{\mu}\rangle\right)^{2}}{I_{E}},
\label{CRB_x_variance}
\end{eqnarray}
meaning that the spatial variance of the electron position along $\hat{\boldsymbol\mu}$-direction $\langle (x^{\mu})^{2}\rangle-\langle x^{\mu}\rangle^{2}$ at a specific energy $E$ has a lower bound set by the energy derivative of the average position $\partial_{E}\langle x^{\mu}\rangle$ and the FIM $I_{E}$.

We now comment on the significance of Eq.~(\ref{CRB_x_variance}). Consider first a homogeneous system without disorder. Because the spatial variance $\langle (x^{\mu})^{2}\rangle-\langle x^{\mu}\rangle^{2}$ is independent of the choice of origin $x^{\mu}=0$ within the measured region, so we may as well set the origin to be at the center of the region. Then it is easy to see that $\langle x^{\mu}\rangle=\sum_{\bf x}x^{\mu}\overline{\rho}({\bf x},E)=0$ at any energy because $x^{\mu}$ is odd and $\overline{\rho}({\bf x},E)$ is even in ${\bf x}$, and this is true at any energy $E$ so $\partial_{E}\langle x^{\mu}\rangle=0$. The CRB in Eq.~(\ref{CRB_x_variance}) then simply states that the spatial variance of electrons in a homogeneous system is positive $\langle (x^{\mu})^{2}\rangle-\langle x^{\mu}\rangle^{2}\geq 0$, a fairly trivial statement. However, in a disordered system $\partial_{E}\langle x^{\mu}\rangle\neq 0$, so the right hand side of Eq.~(\ref{CRB_x_variance}) is no longer zero but a positive number, and hence the lower bound on the spatial variance of electrons is actually increased. In other words, the disorder increases the lower bound on the spatial variance, i.e., the electron position must fluctuate more than $(\partial_{E}\langle x^{\mu}\rangle)^{2}/I_{E}$ in disordered systems.

\begin{figure}[ht]
\begin{center}
\includegraphics[clip=true,width=0.99\columnwidth]{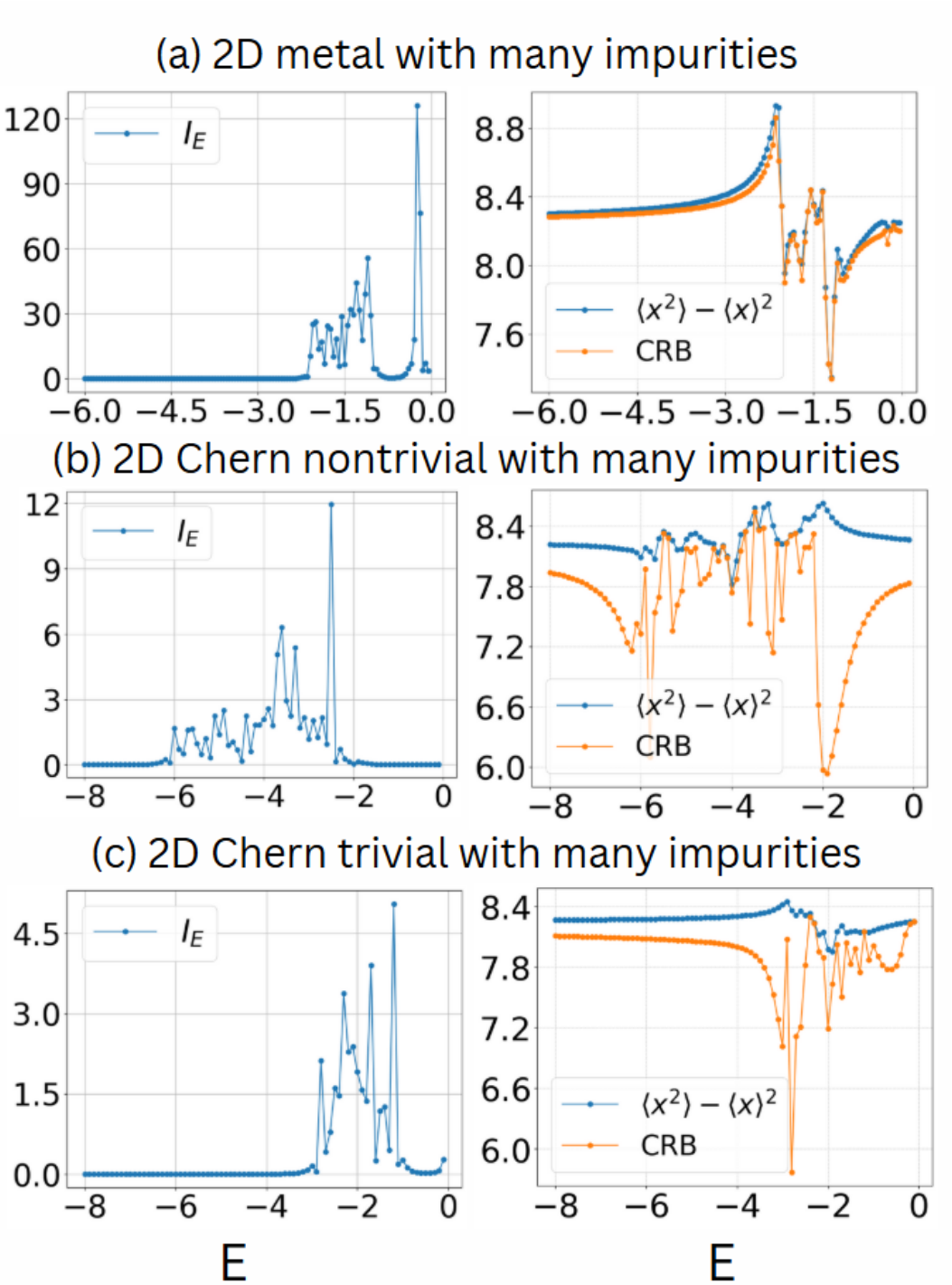}
\caption{Energy space information geometry for disordered (a) 2D metal, 2D Chern insulator in the (b) topologically nontrivial phase and (c) trivial phase, each calculated for one multiple-impurity configuration. For each case, we present the FIM $I_{E}$, the position variance $\langle x^{2}\rangle-\langle x\rangle^{2}$, and the CRB defined as $\langle x^2\rangle-\langle x\rangle^{2}- 
\left(\partial_{E}\langle x\rangle\right)^{2}/I_{E}$, all of which should be positive at any energy regardless of the system parameters.} 
\label{fig:energy_space_CRB_figure}
\end{center}
\end{figure}

\subsection{Applications}

We proceed to use the 2D metal, Chern insulator in the topologically nontrivial and trivial phases introduced in Sec.~\ref{sec:applications_2D_materials} to elaborate the energy space information geometry, using exactly the same parameters. In Figure \ref{fig:energy_space_CRB_figure}, we present the FIM $I_{E}$, position variance along the ${\bf x}$-direction $\langle x^{2}\rangle-\langle x\rangle^{2}$, and the CRB defined as $\langle x^2\rangle-\langle x\rangle^{2}- 
\left(\partial_{E}\langle x\rangle\right)^{2}/I_{E}$, which are plotted as functions of energy $E$. All these quantities are positive at any energy $E$, and indicate that the disorder induces $I_{E}$ and curves the energy space. The two topological phases of Chern insulator display roughly the same magnitude of position variance $\langle x^{2}\rangle-\langle x\rangle^{2}$ and CRB. We conclude that the disorder causes the position variance $\langle x^2\rangle-\langle x\rangle^{2}$ in a way that it is bounded by $\left(\partial_{E}\langle x\rangle\right)^{2}/I_{E}$ at any given energy $E$ where the LDOS is nonzero, indicating an enlarged variance of electron position by disorder as argued above.

\section{Conclusions}

In summary, we propose a formalism to bring the notion of information geometry to solids based on LDOS. By treating the normalized LDOS as a probability density, one can introduce FIMs to describe the distortion of the real space Euclidean manifold or the energy space caused by disorder, in a way that is completely measurable by STM. The associated CRBs set a lower bound on the energy variance $\langle E^{2}\rangle-\langle E\rangle^{2}$ of electrons at a specific position, and on the position variance $\langle (x^{\mu})^{2}\rangle-\langle x^{\mu}\rangle^{2}$ of electrons at a specific energy, both point to new types of information that one can extract from the STM measurement. Furthermore, our formalism is not limited to specific types of solids, be they metallic, insulating, or semiconducting, as demonstrated using 2D metals and topological materials. We anticipate that our formalism can be widely applied to analyzing the deformation of real space and energy space caused by disorder, as well as the lower bound on the energy variance and position variance of electrons in disordered systems, and should be applicable to all different kinds of lattice structures and any types of long or short range defects. The vast arena of these disordered systems await to be further explored.




\acknowledgements

We acknowledge the financial suppoort from the fellowship for productivity in research from CNPq.

\bibliography{Literatur}

\begin{thebibliography}{38}%
\makeatletter
\providecommand \@ifxundefined [1]{%
 \@ifx{#1\undefined}
}%
\providecommand \@ifnum [1]{%
 \ifnum #1\expandafter \@firstoftwo
 \else \expandafter \@secondoftwo
 \fi
}%
\providecommand \@ifx [1]{%
 \ifx #1\expandafter \@firstoftwo
 \else \expandafter \@secondoftwo
 \fi
}%
\providecommand \natexlab [1]{#1}%
\providecommand \enquote  [1]{``#1''}%
\providecommand \bibnamefont  [1]{#1}%
\providecommand \bibfnamefont [1]{#1}%
\providecommand \citenamefont [1]{#1}%
\providecommand \href@noop [0]{\@secondoftwo}%
\providecommand \href [0]{\begingroup \@sanitize@url \@href}%
\providecommand \@href[1]{\@@startlink{#1}\@@href}%
\providecommand \@@href[1]{\endgroup#1\@@endlink}%
\providecommand \@sanitize@url [0]{\catcode `\\12\catcode `\$12\catcode
  `\&12\catcode `\#12\catcode `\^12\catcode `\_12\catcode `\%12\relax}%
\providecommand \@@startlink[1]{}%
\providecommand \@@endlink[0]{}%
\providecommand \url  [0]{\begingroup\@sanitize@url \@url }%
\providecommand \@url [1]{\endgroup\@href {#1}{\urlprefix }}%
\providecommand \urlprefix  [0]{URL }%
\providecommand \Eprint [0]{\href }%
\providecommand \doibase [0]{http://dx.doi.org/}%
\providecommand \selectlanguage [0]{\@gobble}%
\providecommand \bibinfo  [0]{\@secondoftwo}%
\providecommand \bibfield  [0]{\@secondoftwo}%
\providecommand \translation [1]{[#1]}%
\providecommand \BibitemOpen [0]{}%
\providecommand \bibitemStop [0]{}%
\providecommand \bibitemNoStop [0]{.\EOS\space}%
\providecommand \EOS [0]{\spacefactor3000\relax}%
\providecommand \BibitemShut  [1]{\csname bibitem#1\endcsname}%
\let\auto@bib@innerbib\@empty
\bibitem [{\citenamefont {Bourgoin}\ and\ \citenamefont
  {Lannoo}(1983)}]{Bourgoin83}%
  \BibitemOpen
  \bibfield  {author} {\bibinfo {author} {\bibfnamefont {J.}~\bibnamefont
  {Bourgoin}}\ and\ \bibinfo {author} {\bibfnamefont {M.}~\bibnamefont
  {Lannoo}},\ }\href@noop {} {\emph {\bibinfo {title} {Point Defects in
  Semiconductors II: Experimental Aspects}}}\ (\bibinfo  {publisher} {Springer,
  Berlin},\ \bibinfo {year} {1983})\BibitemShut {NoStop}%
\bibitem [{\citenamefont {Queisser}\ and\ \citenamefont
  {Haller}(1998)}]{Queisser98}%
  \BibitemOpen
  \bibfield  {author} {\bibinfo {author} {\bibfnamefont {H.~J.}\ \bibnamefont
  {Queisser}}\ and\ \bibinfo {author} {\bibfnamefont {E.~E.}\ \bibnamefont
  {Haller}},\ }\href {\doibase 10.1126/science.281.5379.945} {\bibfield
  {journal} {\bibinfo  {journal} {Science}\ }\textbf {\bibinfo {volume}
  {281}},\ \bibinfo {pages} {945} (\bibinfo {year} {1998})}\BibitemShut
  {NoStop}%
\bibitem [{\citenamefont {Alkauskas}\ \emph {et~al.}(2016)\citenamefont
  {Alkauskas}, \citenamefont {McCluskey},\ and\ \citenamefont {Van~de
  Walle}}]{Alkauskas16}%
  \BibitemOpen
  \bibfield  {author} {\bibinfo {author} {\bibfnamefont {A.}~\bibnamefont
  {Alkauskas}}, \bibinfo {author} {\bibfnamefont {M.~D.}\ \bibnamefont
  {McCluskey}}, \ and\ \bibinfo {author} {\bibfnamefont {C.~G.}\ \bibnamefont
  {Van~de Walle}},\ }\href {\doibase 10.1063/1.4948245} {\bibfield  {journal}
  {\bibinfo  {journal} {J. Appl. Phys.}\ }\textbf {\bibinfo {volume} {119}},\
  \bibinfo {pages} {181101} (\bibinfo {year} {2016})}\BibitemShut {NoStop}%
\bibitem [{\citenamefont {Friedel}(1952)}]{Friedel52}%
  \BibitemOpen
  \bibfield  {author} {\bibinfo {author} {\bibfnamefont {J.}~\bibnamefont
  {Friedel}},\ }\href {\doibase 10.1080/14786440208561086} {\bibfield
  {journal} {\bibinfo  {journal} {London Edinburgh Philos. Mag. \& J. Sci.}\
  }\textbf {\bibinfo {volume} {43}},\ \bibinfo {pages} {153} (\bibinfo {year}
  {1952})}\BibitemShut {NoStop}%
\bibitem [{\citenamefont {Binnig}\ \emph
  {et~al.}(1982{\natexlab{a}})\citenamefont {Binnig}, \citenamefont {Rohrer},
  \citenamefont {Gerber},\ and\ \citenamefont {Weibel}}]{Binnig82}%
  \BibitemOpen
  \bibfield  {author} {\bibinfo {author} {\bibfnamefont {G.}~\bibnamefont
  {Binnig}}, \bibinfo {author} {\bibfnamefont {H.}~\bibnamefont {Rohrer}},
  \bibinfo {author} {\bibfnamefont {C.}~\bibnamefont {Gerber}}, \ and\ \bibinfo
  {author} {\bibfnamefont {E.}~\bibnamefont {Weibel}},\ }\href {\doibase
  10.1103/PhysRevLett.49.57} {\bibfield  {journal} {\bibinfo  {journal} {Phys.
  Rev. Lett.}\ }\textbf {\bibinfo {volume} {49}},\ \bibinfo {pages} {57}
  (\bibinfo {year} {1982}{\natexlab{a}})}\BibitemShut {NoStop}%
\bibitem [{\citenamefont {Binnig}\ \emph
  {et~al.}(1982{\natexlab{b}})\citenamefont {Binnig}, \citenamefont {Rohrer},
  \citenamefont {Gerber},\ and\ \citenamefont {Weibel}}]{Binnig82_2}%
  \BibitemOpen
  \bibfield  {author} {\bibinfo {author} {\bibfnamefont {G.}~\bibnamefont
  {Binnig}}, \bibinfo {author} {\bibfnamefont {H.}~\bibnamefont {Rohrer}},
  \bibinfo {author} {\bibfnamefont {C.}~\bibnamefont {Gerber}}, \ and\ \bibinfo
  {author} {\bibfnamefont {E.}~\bibnamefont {Weibel}},\ }\href {\doibase
  10.1063/1.92999} {\bibfield  {journal} {\bibinfo  {journal} {Appl. Phys.
  Lett.}\ }\textbf {\bibinfo {volume} {40}},\ \bibinfo {pages} {178} (\bibinfo
  {year} {1982}{\natexlab{b}})}\BibitemShut {NoStop}%
\bibitem [{\citenamefont {Binnig}\ and\ \citenamefont
  {Rohrer}(1987)}]{Binnig87}%
  \BibitemOpen
  \bibfield  {author} {\bibinfo {author} {\bibfnamefont {G.}~\bibnamefont
  {Binnig}}\ and\ \bibinfo {author} {\bibfnamefont {H.}~\bibnamefont
  {Rohrer}},\ }\href {\doibase 10.1103/RevModPhys.59.615} {\bibfield  {journal}
  {\bibinfo  {journal} {Rev. Mod. Phys.}\ }\textbf {\bibinfo {volume} {59}},\
  \bibinfo {pages} {615} (\bibinfo {year} {1987})}\BibitemShut {NoStop}%
\bibitem [{\citenamefont {Hansma}\ and\ \citenamefont
  {Tersoff}(1987)}]{Hansma87}%
  \BibitemOpen
  \bibfield  {author} {\bibinfo {author} {\bibfnamefont {P.~K.}\ \bibnamefont
  {Hansma}}\ and\ \bibinfo {author} {\bibfnamefont {J.}~\bibnamefont
  {Tersoff}},\ }\href {\doibase 10.1063/1.338189} {\bibfield  {journal}
  {\bibinfo  {journal} {J. Appl. Phys.}\ }\textbf {\bibinfo {volume} {61}},\
  \bibinfo {pages} {R1} (\bibinfo {year} {1987})}\BibitemShut {NoStop}%
\bibitem [{\citenamefont {Kubby}\ and\ \citenamefont {Boland}(1996)}]{Kubby96}%
  \BibitemOpen
  \bibfield  {author} {\bibinfo {author} {\bibfnamefont {J.}~\bibnamefont
  {Kubby}}\ and\ \bibinfo {author} {\bibfnamefont {J.}~\bibnamefont {Boland}},\
  }\href {\doibase https://doi.org/10.1016/S0167-5729(97)80001-5} {\bibfield
  {journal} {\bibinfo  {journal} {Surf. Sci. Rep.}\ }\textbf {\bibinfo {volume}
  {26}},\ \bibinfo {pages} {61} (\bibinfo {year} {1996})}\BibitemShut {NoStop}%
\bibitem [{\citenamefont {Yazdani}\ \emph {et~al.}(1999)\citenamefont
  {Yazdani}, \citenamefont {Howald}, \citenamefont {Lutz}, \citenamefont
  {Kapitulnik},\ and\ \citenamefont {Eigler}}]{Yazdani99}%
  \BibitemOpen
  \bibfield  {author} {\bibinfo {author} {\bibfnamefont {A.}~\bibnamefont
  {Yazdani}}, \bibinfo {author} {\bibfnamefont {C.~M.}\ \bibnamefont {Howald}},
  \bibinfo {author} {\bibfnamefont {C.~P.}\ \bibnamefont {Lutz}}, \bibinfo
  {author} {\bibfnamefont {A.}~\bibnamefont {Kapitulnik}}, \ and\ \bibinfo
  {author} {\bibfnamefont {D.~M.}\ \bibnamefont {Eigler}},\ }\href {\doibase
  10.1103/PhysRevLett.83.176} {\bibfield  {journal} {\bibinfo  {journal} {Phys.
  Rev. Lett.}\ }\textbf {\bibinfo {volume} {83}},\ \bibinfo {pages} {176}
  (\bibinfo {year} {1999})}\BibitemShut {NoStop}%
\bibitem [{\citenamefont {Balatsky}\ \emph {et~al.}(2006)\citenamefont
  {Balatsky}, \citenamefont {Vekhter},\ and\ \citenamefont {Zhu}}]{Balatsky06}%
  \BibitemOpen
  \bibfield  {author} {\bibinfo {author} {\bibfnamefont {A.~V.}\ \bibnamefont
  {Balatsky}}, \bibinfo {author} {\bibfnamefont {I.}~\bibnamefont {Vekhter}}, \
  and\ \bibinfo {author} {\bibfnamefont {J.-X.}\ \bibnamefont {Zhu}},\ }\href
  {\doibase 10.1103/RevModPhys.78.373} {\bibfield  {journal} {\bibinfo
  {journal} {Rev. Mod. Phys.}\ }\textbf {\bibinfo {volume} {78}},\ \bibinfo
  {pages} {373} (\bibinfo {year} {2006})}\BibitemShut {NoStop}%
\bibitem [{\citenamefont {Hoffman}\ \emph {et~al.}(2002)\citenamefont
  {Hoffman}, \citenamefont {McElroy}, \citenamefont {Lee}, \citenamefont
  {Lang}, \citenamefont {Eisaki}, \citenamefont {Uchida},\ and\ \citenamefont
  {Davis}}]{Hoffman02}%
  \BibitemOpen
  \bibfield  {author} {\bibinfo {author} {\bibfnamefont {J.~E.}\ \bibnamefont
  {Hoffman}}, \bibinfo {author} {\bibfnamefont {K.}~\bibnamefont {McElroy}},
  \bibinfo {author} {\bibfnamefont {D.-H.}\ \bibnamefont {Lee}}, \bibinfo
  {author} {\bibfnamefont {K.~M.}\ \bibnamefont {Lang}}, \bibinfo {author}
  {\bibfnamefont {H.}~\bibnamefont {Eisaki}}, \bibinfo {author} {\bibfnamefont
  {S.}~\bibnamefont {Uchida}}, \ and\ \bibinfo {author} {\bibfnamefont {J.~C.}\
  \bibnamefont {Davis}},\ }\href {\doibase 10.1126/science.1072640} {\bibfield
  {journal} {\bibinfo  {journal} {Science}\ }\textbf {\bibinfo {volume}
  {297}},\ \bibinfo {pages} {1148} (\bibinfo {year} {2002})}\BibitemShut
  {NoStop}%
\bibitem [{\citenamefont {Fisher}(1925)}]{Fisher25}%
  \BibitemOpen
  \bibfield  {author} {\bibinfo {author} {\bibfnamefont {R.~A.}\ \bibnamefont
  {Fisher}},\ }\href {\doibase 10.1017/S0305004100009580} {\bibfield  {journal}
  {\bibinfo  {journal} {Math. Proc. Camb. Philos. Soc.}\ }\textbf {\bibinfo
  {volume} {22}},\ \bibinfo {pages} {700–725} (\bibinfo {year}
  {1925})}\BibitemShut {NoStop}%
\bibitem [{\citenamefont {Amari}\ and\ \citenamefont
  {Nagaoka}(2000)}]{Amari00}%
  \BibitemOpen
  \bibfield  {author} {\bibinfo {author} {\bibfnamefont {S.}~\bibnamefont
  {Amari}}\ and\ \bibinfo {author} {\bibfnamefont {H.}~\bibnamefont
  {Nagaoka}},\ }\href@noop {} {\emph {\bibinfo {title} {Methods of Information
  Geometry}}}\ (\bibinfo  {publisher} {American Mathematical Society},\
  \bibinfo {address} {Providence, RI},\ \bibinfo {year} {2000})\ p.\ \bibinfo
  {pages} {206}\BibitemShut {NoStop}%
\bibitem [{\citenamefont {Amari}(2016)}]{Amari16}%
  \BibitemOpen
  \bibfield  {author} {\bibinfo {author} {\bibfnamefont {S.}~\bibnamefont
  {Amari}},\ }\href@noop {} {\emph {\bibinfo {title} {Information Geometry and
  Its Applications}}}\ (\bibinfo  {publisher} {Springer},\ \bibinfo {address}
  {Tokyo},\ \bibinfo {year} {2016})\ p.\ \bibinfo {pages} {374}\BibitemShut
  {NoStop}%
\bibitem [{\citenamefont {Nagy}(2022)}]{Nagy22}%
  \BibitemOpen
  \bibfield  {author} {\bibinfo {author} {\bibfnamefont {A.}~\bibnamefont
  {Nagy}},\ }\href {\doibase https://doi.org/10.1002/qua.26679} {\bibfield
  {journal} {\bibinfo  {journal} {Int. J. Quantum Chem.}\ }\textbf {\bibinfo
  {volume} {122}},\ \bibinfo {pages} {e26679} (\bibinfo {year}
  {2022})}\BibitemShut {NoStop}%
\bibitem [{\citenamefont {Prokopenko}\ \emph {et~al.}(2011)\citenamefont
  {Prokopenko}, \citenamefont {Lizier}, \citenamefont {Obst},\ and\
  \citenamefont {Wang}}]{Prokopenko11}%
  \BibitemOpen
  \bibfield  {author} {\bibinfo {author} {\bibfnamefont {M.}~\bibnamefont
  {Prokopenko}}, \bibinfo {author} {\bibfnamefont {J.~T.}\ \bibnamefont
  {Lizier}}, \bibinfo {author} {\bibfnamefont {O.}~\bibnamefont {Obst}}, \ and\
  \bibinfo {author} {\bibfnamefont {X.~R.}\ \bibnamefont {Wang}},\ }\href
  {\doibase 10.1103/PhysRevE.84.041116} {\bibfield  {journal} {\bibinfo
  {journal} {Phys. Rev. E}\ }\textbf {\bibinfo {volume} {84}},\ \bibinfo
  {pages} {041116} (\bibinfo {year} {2011})}\BibitemShut {NoStop}%
\bibitem [{\citenamefont {Curilef}\ \emph {et~al.}(2005)\citenamefont
  {Curilef}, \citenamefont {Pennini},\ and\ \citenamefont
  {Plastino}}]{Curilef05}%
  \BibitemOpen
  \bibfield  {author} {\bibinfo {author} {\bibfnamefont {S.}~\bibnamefont
  {Curilef}}, \bibinfo {author} {\bibfnamefont {F.}~\bibnamefont {Pennini}}, \
  and\ \bibinfo {author} {\bibfnamefont {A.}~\bibnamefont {Plastino}},\ }\href
  {\doibase 10.1103/PhysRevB.71.024420} {\bibfield  {journal} {\bibinfo
  {journal} {Phys. Rev. B}\ }\textbf {\bibinfo {volume} {71}},\ \bibinfo
  {pages} {024420} (\bibinfo {year} {2005})}\BibitemShut {NoStop}%
\bibitem [{\citenamefont {Cram\'{e}r}(1946)}]{Cramer46}%
  \BibitemOpen
  \bibfield  {author} {\bibinfo {author} {\bibfnamefont {H.}~\bibnamefont
  {Cram\'{e}r}},\ }\href@noop {} {\emph {\bibinfo {title} {Mathematical Methods
  of Statistics}}}\ (\bibinfo  {publisher} {Princeton University Press,
  Princeton},\ \bibinfo {year} {1946})\ p.\ \bibinfo {pages} {575}\BibitemShut
  {NoStop}%
\bibitem [{\citenamefont {Rao}(1945)}]{Rao45}%
  \BibitemOpen
  \bibfield  {author} {\bibinfo {author} {\bibfnamefont {C.~R.}\ \bibnamefont
  {Rao}},\ }\href@noop {} {\bibfield  {journal} {\bibinfo  {journal} {Bull.
  Calcutta Math. Soc.}\ }\textbf {\bibinfo {volume} {37}},\ \bibinfo {pages}
  {81} (\bibinfo {year} {1945})}\BibitemShut {NoStop}%
\bibitem [{\citenamefont {Bardeen}(1961)}]{Bardeen61}%
  \BibitemOpen
  \bibfield  {author} {\bibinfo {author} {\bibfnamefont {J.}~\bibnamefont
  {Bardeen}},\ }\href {\doibase 10.1103/PhysRevLett.6.57} {\bibfield  {journal}
  {\bibinfo  {journal} {Phys. Rev. Lett.}\ }\textbf {\bibinfo {volume} {6}},\
  \bibinfo {pages} {57} (\bibinfo {year} {1961})}\BibitemShut {NoStop}%
\bibitem [{\citenamefont {Tersoff}\ and\ \citenamefont
  {Hamann}(1983)}]{Tersoff83}%
  \BibitemOpen
  \bibfield  {author} {\bibinfo {author} {\bibfnamefont {J.}~\bibnamefont
  {Tersoff}}\ and\ \bibinfo {author} {\bibfnamefont {D.~R.}\ \bibnamefont
  {Hamann}},\ }\href {\doibase 10.1103/PhysRevLett.50.1998} {\bibfield
  {journal} {\bibinfo  {journal} {Phys. Rev. Lett.}\ }\textbf {\bibinfo
  {volume} {50}},\ \bibinfo {pages} {1998} (\bibinfo {year}
  {1983})}\BibitemShut {NoStop}%
\bibitem [{\citenamefont {Tersoff}\ and\ \citenamefont
  {Hamann}(1985)}]{Tersoff85}%
  \BibitemOpen
  \bibfield  {author} {\bibinfo {author} {\bibfnamefont {J.}~\bibnamefont
  {Tersoff}}\ and\ \bibinfo {author} {\bibfnamefont {D.~R.}\ \bibnamefont
  {Hamann}},\ }\href {\doibase 10.1103/PhysRevB.31.805} {\bibfield  {journal}
  {\bibinfo  {journal} {Phys. Rev. B}\ }\textbf {\bibinfo {volume} {31}},\
  \bibinfo {pages} {805} (\bibinfo {year} {1985})}\BibitemShut {NoStop}%
\bibitem [{\citenamefont {Chen}(1990{\natexlab{a}})}]{Chen90}%
  \BibitemOpen
  \bibfield  {author} {\bibinfo {author} {\bibfnamefont {C.~J.}\ \bibnamefont
  {Chen}},\ }\href {\doibase 10.1103/PhysRevLett.65.448} {\bibfield  {journal}
  {\bibinfo  {journal} {Phys. Rev. Lett.}\ }\textbf {\bibinfo {volume} {65}},\
  \bibinfo {pages} {448} (\bibinfo {year} {1990}{\natexlab{a}})}\BibitemShut
  {NoStop}%
\bibitem [{\citenamefont {Chen}(1990{\natexlab{b}})}]{Chen90_2}%
  \BibitemOpen
  \bibfield  {author} {\bibinfo {author} {\bibfnamefont {C.~J.}\ \bibnamefont
  {Chen}},\ }\href {\doibase 10.1103/PhysRevB.42.8841} {\bibfield  {journal}
  {\bibinfo  {journal} {Phys. Rev. B}\ }\textbf {\bibinfo {volume} {42}},\
  \bibinfo {pages} {8841} (\bibinfo {year} {1990}{\natexlab{b}})}\BibitemShut
  {NoStop}%
\bibitem [{\citenamefont {Chen}(2021)}]{Chen21_STM_book}%
  \BibitemOpen
  \bibfield  {author} {\bibinfo {author} {\bibfnamefont {C.~J.}\ \bibnamefont
  {Chen}},\ }\href@noop {} {\emph {\bibinfo {title} {Introduction to Scanning
  Tunneling Microscopy}}},\ \bibinfo {edition} {3rd}\ ed.\ (\bibinfo
  {publisher} {Oxford University Press},\ \bibinfo {address} {Oxford, UK},\
  \bibinfo {year} {2021})\BibitemShut {NoStop}%
\bibitem [{\citenamefont {Oliveira}\ and\ \citenamefont
  {Chen}(2025)}]{Oliveira25_real_space_quantum_metric}%
  \BibitemOpen
  \bibfield  {author} {\bibinfo {author} {\bibfnamefont {L.~A.}\ \bibnamefont
  {Oliveira}}\ and\ \bibinfo {author} {\bibfnamefont {W.}~\bibnamefont
  {Chen}},\ }\href {\doibase 10.1103/PhysRevB.111.094202} {\bibfield  {journal}
  {\bibinfo  {journal} {Phys. Rev. B}\ }\textbf {\bibinfo {volume} {111}},\
  \bibinfo {pages} {094202} (\bibinfo {year} {2025})}\BibitemShut {NoStop}%
\bibitem [{\citenamefont {Provost}\ and\ \citenamefont
  {Vallee}(1980)}]{Provost80}%
  \BibitemOpen
  \bibfield  {author} {\bibinfo {author} {\bibfnamefont {J.~P.}\ \bibnamefont
  {Provost}}\ and\ \bibinfo {author} {\bibfnamefont {G.}~\bibnamefont
  {Vallee}},\ }\href {https://projecteuclid.org:443/euclid.cmp/1103908308}
  {\bibfield  {journal} {\bibinfo  {journal} {Comm. Math. Phys.}\ }\textbf
  {\bibinfo {volume} {76}},\ \bibinfo {pages} {289} (\bibinfo {year}
  {1980})}\BibitemShut {NoStop}%
\bibitem [{\citenamefont {van~den Bos}(2007)}]{vandenBos07}%
  \BibitemOpen
  \bibfield  {author} {\bibinfo {author} {\bibfnamefont {A.}~\bibnamefont
  {van~den Bos}},\ }\href@noop {} {\emph {\bibinfo {title} {Parameter
  Estimation for Scientists and Engineers}}}\ (\bibinfo  {publisher} {John
  Wiley and Sons, Hoboken},\ \bibinfo {year} {2007})\BibitemShut {NoStop}%
\bibitem [{\citenamefont {Andrei}\ \emph {et~al.}(2012)\citenamefont {Andrei},
  \citenamefont {Li},\ and\ \citenamefont {Du}}]{Andrei12}%
  \BibitemOpen
  \bibfield  {author} {\bibinfo {author} {\bibfnamefont {E.~Y.}\ \bibnamefont
  {Andrei}}, \bibinfo {author} {\bibfnamefont {G.}~\bibnamefont {Li}}, \ and\
  \bibinfo {author} {\bibfnamefont {X.}~\bibnamefont {Du}},\ }\href {\doibase
  10.1088/0034-4885/75/5/056501} {\bibfield  {journal} {\bibinfo  {journal}
  {Rep. Prog. Phys.}\ }\textbf {\bibinfo {volume} {75}},\ \bibinfo {pages}
  {056501} (\bibinfo {year} {2012})}\BibitemShut {NoStop}%
\bibitem [{\citenamefont {Chen}(2020)}]{Chen20_absence_edge_current}%
  \BibitemOpen
  \bibfield  {author} {\bibinfo {author} {\bibfnamefont {W.}~\bibnamefont
  {Chen}},\ }\href {\doibase 10.1103/PhysRevB.101.195120} {\bibfield  {journal}
  {\bibinfo  {journal} {Phys. Rev. B}\ }\textbf {\bibinfo {volume} {101}},\
  \bibinfo {pages} {195120} (\bibinfo {year} {2020})}\BibitemShut {NoStop}%
\bibitem [{\citenamefont {Molignini}\ \emph {et~al.}(2023)\citenamefont
  {Molignini}, \citenamefont {Lapierre}, \citenamefont {Chitra},\ and\
  \citenamefont {Chen}}]{Molignini23_Chern_marker}%
  \BibitemOpen
  \bibfield  {author} {\bibinfo {author} {\bibfnamefont {P.}~\bibnamefont
  {Molignini}}, \bibinfo {author} {\bibfnamefont {B.}~\bibnamefont {Lapierre}},
  \bibinfo {author} {\bibfnamefont {R.}~\bibnamefont {Chitra}}, \ and\ \bibinfo
  {author} {\bibfnamefont {W.}~\bibnamefont {Chen}},\ }\href {\doibase
  10.21468/SciPostPhysCore.6.3.059} {\bibfield  {journal} {\bibinfo  {journal}
  {SciPost Phys. Core}\ }\textbf {\bibinfo {volume} {6}},\ \bibinfo {pages}
  {059} (\bibinfo {year} {2023})}\BibitemShut {NoStop}%
\bibitem [{\citenamefont {Bianco}\ and\ \citenamefont
  {Resta}(2011)}]{Bianco11}%
  \BibitemOpen
  \bibfield  {author} {\bibinfo {author} {\bibfnamefont {R.}~\bibnamefont
  {Bianco}}\ and\ \bibinfo {author} {\bibfnamefont {R.}~\bibnamefont {Resta}},\
  }\href {\doibase 10.1103/PhysRevB.84.241106} {\bibfield  {journal} {\bibinfo
  {journal} {Phys. Rev. B}\ }\textbf {\bibinfo {volume} {84}},\ \bibinfo
  {pages} {241106} (\bibinfo {year} {2011})}\BibitemShut {NoStop}%
\bibitem [{\citenamefont {Prodan}\ \emph {et~al.}(2010)\citenamefont {Prodan},
  \citenamefont {Hughes},\ and\ \citenamefont {Bernevig}}]{Prodan10}%
  \BibitemOpen
  \bibfield  {author} {\bibinfo {author} {\bibfnamefont {E.}~\bibnamefont
  {Prodan}}, \bibinfo {author} {\bibfnamefont {T.~L.}\ \bibnamefont {Hughes}},
  \ and\ \bibinfo {author} {\bibfnamefont {B.~A.}\ \bibnamefont {Bernevig}},\
  }\href {\doibase 10.1103/PhysRevLett.105.115501} {\bibfield  {journal}
  {\bibinfo  {journal} {Phys. Rev. Lett.}\ }\textbf {\bibinfo {volume} {105}},\
  \bibinfo {pages} {115501} (\bibinfo {year} {2010})}\BibitemShut {NoStop}%
\bibitem [{\citenamefont {Costa}\ \emph {et~al.}(2019)\citenamefont {Costa},
  \citenamefont {Schleder}, \citenamefont {Buongiorno~Nardelli}, \citenamefont
  {Lewenkopf},\ and\ \citenamefont {Fazzio}}]{Costa19}%
  \BibitemOpen
  \bibfield  {author} {\bibinfo {author} {\bibfnamefont {M.}~\bibnamefont
  {Costa}}, \bibinfo {author} {\bibfnamefont {G.~R.}\ \bibnamefont {Schleder}},
  \bibinfo {author} {\bibfnamefont {M.}~\bibnamefont {Buongiorno~Nardelli}},
  \bibinfo {author} {\bibfnamefont {C.}~\bibnamefont {Lewenkopf}}, \ and\
  \bibinfo {author} {\bibfnamefont {A.}~\bibnamefont {Fazzio}},\ }\href
  {\doibase 10.1021/acs.nanolett.9b03881} {\bibfield  {journal} {\bibinfo
  {journal} {Nano Lett.}\ }\textbf {\bibinfo {volume} {19}},\ \bibinfo {pages}
  {8941} (\bibinfo {year} {2019})}\BibitemShut {NoStop}%
\bibitem [{\citenamefont {Ul\ifmmode~\check{c}\else \v{c}\fi{}akar}\ \emph
  {et~al.}(2020)\citenamefont {Ul\ifmmode~\check{c}\else \v{c}\fi{}akar},
  \citenamefont {Mravlje},\ and\ \citenamefont {Rejec}}]{Ulcakar20}%
  \BibitemOpen
  \bibfield  {author} {\bibinfo {author} {\bibfnamefont {L.}~\bibnamefont
  {Ul\ifmmode~\check{c}\else \v{c}\fi{}akar}}, \bibinfo {author} {\bibfnamefont
  {J.}~\bibnamefont {Mravlje}}, \ and\ \bibinfo {author} {\bibfnamefont
  {T.~c.~v.}\ \bibnamefont {Rejec}},\ }\href {\doibase
  10.1103/PhysRevLett.125.216601} {\bibfield  {journal} {\bibinfo  {journal}
  {Phys. Rev. Lett.}\ }\textbf {\bibinfo {volume} {125}},\ \bibinfo {pages}
  {216601} (\bibinfo {year} {2020})}\BibitemShut {NoStop}%
\bibitem [{\citenamefont {d'Ornellas}\ \emph {et~al.}(2022)\citenamefont
  {d'Ornellas}, \citenamefont {Barnett},\ and\ \citenamefont
  {Lee}}]{dOrnellas22}%
  \BibitemOpen
  \bibfield  {author} {\bibinfo {author} {\bibfnamefont {P.}~\bibnamefont
  {d'Ornellas}}, \bibinfo {author} {\bibfnamefont {R.}~\bibnamefont {Barnett}},
  \ and\ \bibinfo {author} {\bibfnamefont {D.~K.~K.}\ \bibnamefont {Lee}},\
  }\href {\doibase 10.1103/PhysRevB.106.155124} {\bibfield  {journal} {\bibinfo
   {journal} {Phys. Rev. B}\ }\textbf {\bibinfo {volume} {106}},\ \bibinfo
  {pages} {155124} (\bibinfo {year} {2022})}\BibitemShut {NoStop}%
\bibitem [{\citenamefont {Oliveira}\ and\ \citenamefont
  {Chen}(2024)}]{Oliveira24_impurity_marker}%
  \BibitemOpen
  \bibfield  {author} {\bibinfo {author} {\bibfnamefont {L.~A.}\ \bibnamefont
  {Oliveira}}\ and\ \bibinfo {author} {\bibfnamefont {W.}~\bibnamefont
  {Chen}},\ }\href {\doibase 10.1103/PhysRevB.109.094202} {\bibfield  {journal}
  {\bibinfo  {journal} {Phys. Rev. B}\ }\textbf {\bibinfo {volume} {109}},\
  \bibinfo {pages} {094202} (\bibinfo {year} {2024})}\BibitemShut {NoStop}%
\end{thebibliography}%

\end{document}